\documentclass[11pt]{article}
\usepackage{acl2015}
\usepackage{times}
\usepackage{url}
\usepackage{xcolor}
\usepackage{soul} 
\usepackage{amsmath,amssymb,amsfonts}
\usepackage{graphicx}
\usepackage{xparse,array,booktabs}
\usepackage{makecell}
\usepackage{tikz}
\usepackage{pgfplots}
\usepackage{pgfplotstable}
\usepgfplotslibrary{dateplot}

\pgfplotsset{width=7cm,compat=1.8}
\usepackage{sfmath}

\title{ Discovering Affinity Relationships between Personality Types}

\author{ Jean Marie Tshimula,$^{1}$ Belkacem Chikhaoui,$^{1,2}$ Shengrui Wang$^{1}$ \\
  \normalsize $^{1}$Department of Computer Science, Université de Sherbrooke, QC J1K 2R1, Canada \\
  $^{2}$LICEF Research Center, Université TÉLUQ, QC H2S 3L5, Canada\\
 {\tt \{kabj2801,shengrui.wang\}@usherbrooke.ca } \\
 {\tt {belkacem.chikhaoui}@teluq.ca }
 }
 
\date{}

\begin{document}
\maketitle
\begin{abstract}
Psychology research findings suggest that personality is related to differences in friendship characteristics and that some personality traits correlate with linguistic behavior. In this paper, we investigate the influence that personality may have on affinity formation. To this end, we derive affinity relationships from social media interactions, examine personality based on language use to discover the emotional stability of affinity relationships, and measure semantic similarity at the personality type level to understand the logic behind the development of affinity. Specifically, we conduct extensive experiments using a publicly available dataset containing information on individuals who self-identified with a Myers-Briggs personality type. Our results identify certain influential personality types that weigh more heavily on affinity relationships and show that personality can be predicted from spontaneous language with an F-1 score superior to 0.76. Future research avenues are proposed.
\end{abstract}

\section{Introduction}
The study of friendship has long been a mainstay of research on developmental psychology \cite{paX,paY}. There are various stages of friendship, including formation, maintenance, and dissolution. Our focus here is on friendship formation. To some extent, the process of friendship formation can be fairly similar in real-life and online social networks, in that it involves the transition from strangers to acquaintances to friends. Individuals engage in interactions to get to know each other and forge the affective bond that characterizes a friendship. While friendships are formed differently and for various reasons, all friendships undergo a formation process. Research has shown that the formation process may be influenced by various factors, which may be environmental, individual, situational or dyadic, such as personality similarity effects \cite{pa04}. Intuitively, people who share common values, tenets, convictions, and personality traits are more likely to become friends. Research on personality and friendship has yielded profound discoveries, but the two are usually studied singly; their interdependence has been investigated only recently \cite{pa03,pa04,paY}. This paper attempts to bridge the gap between personality and friendship by utilizing online social interactions to investigate the psychological processes underlying the development of affinity.

Our paper specifically regards affinity in friendships. To the best of our knowledge, our work is the first to address the combination of affinity and personality. Practically, investigating affinity and personality is of interest not only for psychology but also for commercial applications, including in mental health services to understand the psychological aspects and the effects of mental illness on individual patients and social systems. The combination of affinity and personality allows us to understand how individuals with similar personality traits get to develop their affinity and discern what attracts an individual to another.  

Most studies on personality use questionnaires (and/or written essays) to assess personal behavioral preferences. This approach inherently inhibits the expression of individual traits and makes it difficult to track language use and interactions between subjects. To efficiently conduct an analysis of language use between individuals based on their personality types requires that the data be annotated beforehand. The lack of labeled data impedes the potential of computational personality recognition to yield reliable, high-quality results \cite{pa00Now}. It should be noted that manually annotated datasets are expensive and hard to obtain. To overcome the limitation of the small size of annotated data samples and closed-vocabulary, we chose to utilize social media data \cite{paPlankHorvy}. Specifically, we have collected a corpus of 758,426 English tweets with self-assessed Myers-Briggs Type Indicators \cite{paBriggs}, denoted MBTI. {\color{black}The MBTI assessment is based on research and personalized preferences and can contribute important information to the understanding of individual psychological functions such as intuition, sensation, thinking, feeling, etc.} The MBTI model defines four binary dimensions—Introversion–Extraversion (I-E), Intuition–Sensing (N-S), Feeling–Thinking (F-T), Perception–Judgment (P-J)—that combine to yield 16 personality types into which individuals may be classified: e.g., INFP, ESTJ, ISFJ, etc. The characteristics of each MBTI personality type are described in Table \ref{MBTICharacteristicsAffPersXX2eXOiP}.

Furnham et al.~\shortcite{paFurnham} performed a correlation analysis of personality traits between the MBTI and Big Five models and showed that the Big Five dimension Extraversion correlates with the MBTI (I-E), Openness to Experience correlates with (N-S), Agreeableness with (F-T), and Conscientiousness with (P-J). The rationale for using the MBTI model is that it facilitates the collection of gold-standard labeled data compared with the Big Five. The 16 MBTI personality types are simple to manipulate to account for personality differences. Since the MBTI model lacks reference to the Big Five Neuroticism dimension, we investigate the language use of individuals who self-identified with an MBTI personality type in order to discover their emotional stability. To this end, we use a psychometrically validated system to extract emotion-based psycholinguistic features. We utilize self-identified MBTI personality types as annotations and train five different models to predict personality on a linguistic level. In order to understand the factors that contribute to the establishment of affinities, we investigate emotional stability and semantic similarity in affinity pairings based on their personality types. We seek to identify the influential personality types that weigh more heavily on affinity relationships.

To summarize, we make the following contributions:

\begin{enumerate}
    \item We show the effectiveness of our data collection and data pre-processing strategy to gather social media postings containing MBTI personality types.
    \item We discover personality-based affinity relationships from social media interactions and investigate the emotional stability of affinity relationships based on language use.
    \item We measure semantic similarity in affinity pairings at the personality level to understand the logic behind the development of affinity.
    \item We propose an approach to detect the influence that personality has on affinity formation.
\end{enumerate}

In line with these contributions, the remainder of this paper is organized as follows. Section \ref{PrerWERelatedWorkAffPErson} discusses some related work. Section \ref{PrsAffDtsHvSkO35ZWp82n} describes the strategies utilized to extract and process our dataset. In Section \ref{PrsAffMthdUiO12cX017345E}, we explain our methodology, from affinity computation, through the formulation of affinity graphs with personality traits, to detect the influence of personality on affinity. We then present our experimental setup and discuss results on similarity, psycholinguistic features, and prediction in Section \ref{PrsAffExp43tGhZcqk73}. In all cases, the results we obtain are thoroughly analyzed. Results are extensively discussed in Section \ref{PrDscss9189erhd091ue}. Finally, Section \ref{PrsAffClsO73ByZol} puts forward some concluding remarks and presents future directions.

\section{Related Work}\label{PrerWERelatedWorkAffPErson}

Most studies on personality and friendship rely on the most popular personality construct in contemporary psychology, the Big Five personality traits \cite{pa01}, to scrutinize interpersonal attraction \cite{pa08,pa88} and psychological well-being (satisfaction, happiness, self-acceptance, etc). Demir and Weitekamp~\shortcite{pa05} investigated the role that friendship plays in happiness and showed that friendship quality can contribute to happiness above and beyond the influence of gender and personality. Laakasuo et al.~\shortcite{pa03} and Wilson et al.~\shortcite{pa02} have focused on similarities between friends and friendship patterns and found that certain personality traits are important predictors of friendship satisfaction. For instance, people who exhibit the personality traits of extroversion, agreeableness, and conscientiousness have more satisfying relationships than those who rank high in the personality trait of neuroticism. Neurotic people are linked to lower satisfaction. This may be partly explained by the fact that emotionally unstable people can be somewhat on the dramatic or high-maintenance side. Additionally, studies have shown that conscientious people have fewer unemployed friends and are more likely to have friends of the same gender, while people with high openness to experience are more likely to befriend those of different gender and ethnicity \cite{pa03}. Openness to experience seems to be associated with exploratory and complementary friendship styles, while agreeableness and a lesser degree of extroversion are related to more traditional friendship ties, stressing stability and proximity of friends \cite{pa03}. Extroversion, conscientiousness, and openness to experience have all been shown to influence relationship development, but their effects are inconsistent \cite{pa04}.

Understanding the factors that contribute to interpersonal attraction and lead to friendships can be of crucial importance. Roberts-Griffin~\shortcite{pa88} consequently focused on three factors (namely propinquity effect, similarity, and attractiveness) and found that these factors have a significant effect on whom individuals befriend.~The three factors can be important when selecting close friends. Furthermore, Roberts-Griffin~\shortcite{pa88} asserted that these factors can also work in negative ways: that is, individuals can come to dislike others in the presence of these three factors.

\begin{table*}[ht]
\fontsize{9pt}{9pt}\selectfont
\setlength\tabcolsep{2.4pt}
\centering
\caption{Data summary and distribution. We collected Twitter data self-identified with MBTI personality types and calculated the percentage of each type in the dataset. We observe that INFJ comprises a large amount of data, and ISTP a much smaller amount.}
\label{TblOne45XTiOH2vPq1}
\begin{tabular}{ c c c c c c c c c c c c c c c c c}
\toprule
\text{Type} & \text{ISTJ} & \text{ISFP} & \text{INFP} & \text{ESFJ} & \text{ISTP} & \text{ISFJ} & \text{INFJ} & \text{ENTP} & \text{INTP} & \text{INTJ} & \text{ESFP} & \text{ENTJ} & \text{ESTP} & \text{ESTJ} & \text{ENFP} & \text{ENFJ} \\
\toprule
\text{\%} & {10.3} & {6.2} & {7.3} & {8} & {1.7} & {9.2} & {12.3} & {2.6} & {3.3} & {8.7} & {3.5} & {5.6} & {2.9} & {6.5} & {6.8} & {5.1}\\
\bottomrule
\end{tabular}
\end{table*}

Friendships in social media are generally inferred from structural features \cite{paAnts,paTang}. However, relying solely on structural features may fail to extract some essential friendship character traits. For instance, in online social interactions, individuals may appear to be closer to one another based on social network structure, while they do not always show mutual appreciation and their interactions entail some divergent opinions. Tshimula et al.~\shortcite{paAff2020} therefore combined structural features and the content of interactions between individuals to understand their friendships and measure affinity scores between them, and predicted affinity relationships arising from the influence of certain individuals. We utilize the approach introduced by \cite{paAff2020} to generate a personality-based affinity graph. We measure emotional stability and semantic similarity between affinity pairings. We then apply graph clustering to discover the connectivity between nodes within each cluster and build a methodology to detect the influence that personality has on affinity. {\color{black}The rationale behind the detection of the influence of personality on affinity within clusters is to identify all possible groups formed by individuals based on their interactions}.

\section{Datasets}\label{PrsAffDtsHvSkO35ZWp82n}

For this research, we prepared a dataset consisting of tweets from individuals who publicly self-identified with one of the 16 MBTI personality types. Specifically, we collected tweets containing any of the 16 MBTI personality types plus the terms ``MBTI'', ``Briggs'' and/or ``Myers''. For privacy and ethical considerations, we avoid displaying personally identifiable information, especially names and pseudonyms. Consequently, we randomly replaced such information to ensure the anonymity and privacy of the data. \\

\noindent
\textbf{Dataset A.}
To process the data, we removed tweets written in a language other than English. We eliminated retweets and all tweets comprising more than one personality type, and removed redundant tweets. We utilized Botometer\footnote{https://botometer.osome.iu.edu/}, a web-based tool that uses machine learning to classify Twitter accounts as bot or human by looking at features such as friends, social network structure, temporal activity, language and sentiment. Botometer yields an overall bot score along with several other scores that provides a measure of the likelihood that the account is a bot. Bot scores display on a 0-to-5 scale with zero being most human-like and five being the most bot-like. We therefore removed arbitrarily all users for which the overall bot score is higher than 2.5. We believe that accounts displaying the score of 2.5 are in the middle of the scale, and these accounts are on a relatively neutral ground. It could be difficult to classify the bot score 2.5 as human or bot. That is the reason why we consider as a bot any account displaying an overall bot score greater than or equal to 2.5. The rationale behind this is to ensure reliable data collection.

In order to thoroughly examine the language use and how it varies across each personality type, we discarded all tweets belonging to the same user in which the MBTI personality types are different. Overall, we extracted 758,426 tweets, for the same number of users. Table \ref{TblOne45XTiOH2vPq1} outlines dataset A and shows the distribution over the MBTI personality types. We report that 9.1\% of this dataset contains mentions, i.e., the @ symbol plus a username.\\

\noindent
\textbf{Dataset B.}
Since the algorithm of affinity relies heavily on mentions between users \cite{paAff2020}, we retrieved the most recent tweets (up to 200) for each self-identified user of dataset A. Specifically, we obtained a total of 25,253,604 tweets with an average of 33 tweets per user. We believe that in these tweets users are more likely to make use of spontaneous language in various contexts to express themselves than when they self-report or talk about their MBTI personality type in a single tweet.\\

\noindent
\textbf{Dataset for MBTI personality type prediction.}
The average number of words in dataset A is 27 per user, while in dataset B, there are 4,843 per user. We therefore took all tweets in dataset B for each user and labeled them with the MBTI personality type. {\color{black}The annotation of dataset B facilitates the extraction of behavioral patterns related to each MBTI personality type to develop a model that can predict personality on each of the 16 MBTI personality types (see Table \ref{tabXVC45aYTDe}).}

\begin{figure*}
    \centering
    \includegraphics[width=2.1\columnwidth]{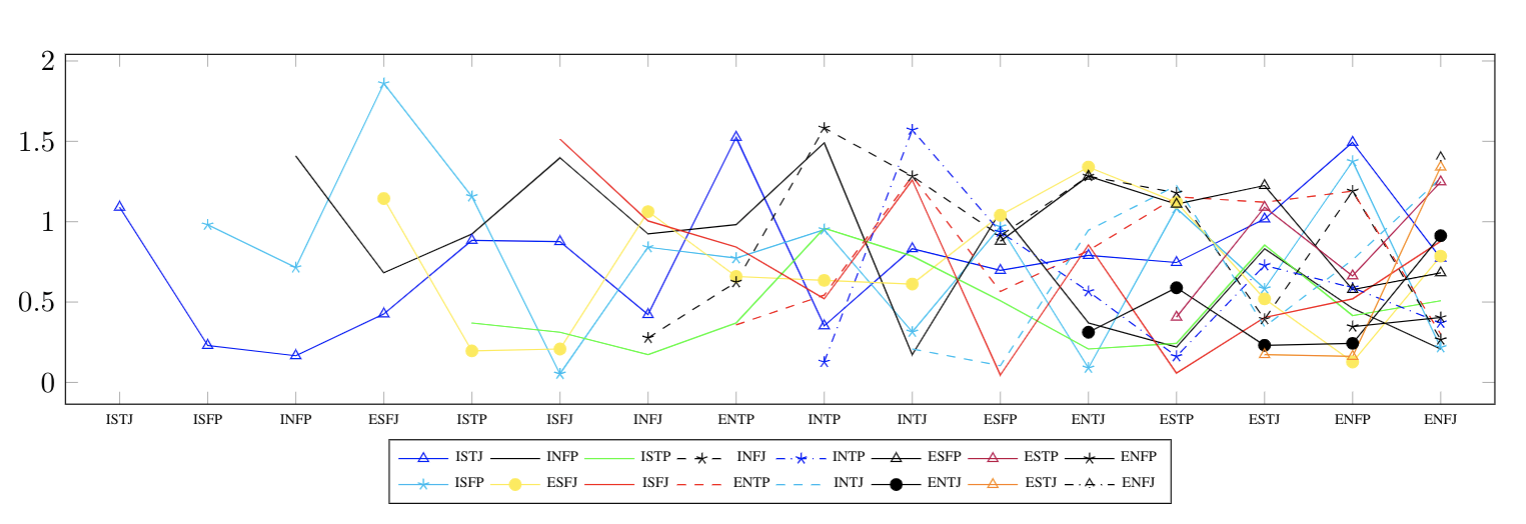}
    \caption{Affinity percentages between the {136} combinations of the {16 MBTI} personality types. {Affinity relationships that achieve a percentage superior to {1.3\%} are:  {ENTP--ISTJ (1.53\%)}, {ENFP-ISTJ (1.49\%)}, {ESFJ--ISFP (1.86\%)}, {ENFP--ISFP (1.38\%)}, {INFP--INFP (1.41\%)}, {ISFJ--INFP (1.4\%)}, {INTP--INFP (1.49\%)}, {ISFJ--ISFJ (1.51\%)}, {INTP--INFJ (1.58\%)}, {INTJ-INTP (1.57\%)}, {ENFJ--ESTJ (1.34\%)} and {ENFJ--ENFJ (1.4\%)}.}}
    \label{FigXDESQIO7865}
\end{figure*}

\section{Methodology}\label{PrsAffMthdUiO12cX017345E}
We take the set of users who publicly self-identified with an MBTI personality type (see dataset A), and verify whether relationships exist between them. To this end, we regard mentions in dataset B to seek to identify tweets that link these users to one another. We obtained an overall of 3,481,737 tweets bearing mentions, that is, 13.8\% of the entire dataset B. We are particularly interested in {\color{black}tracking and investigating} social mentions. The affinity algorithm, HAR-search, utilizes mentions to effectively understand their implications in social interactions, including the sentiment and the context in which mentions were tagged in discussion threads, in order to derive affinity relationships. For a good retrospective and prospective summary of the concept of affinity in social media, we refer the reader to HAR-search \cite{paAff2020}.

\subsection{Affinity Computation}
Affinity relationships can basically be observed from a set of characteristics, including mutual understanding, reciprocal and common interests, sympathy, harmonious communication, and agreement between individuals \cite{paAff2020}. In this paper, we utilize HAR-search to derive affinity scores between users from online discussions. Specifically, HAR-search considers mentions and the flow of discussions to capture minute details and contexts of interactions based on their time-series order. HAR-search extracts affinity-relevant signals from interactions, based on their sentiment and context, and then models these signals in the form of sequences. Markov chain models are then used to quantify these sequences to yield affinity scores. These values denote the degree of affinity between a pair of users. The rationale for using HAR-search is that it facilitates the generation of a Markov transition probability matrix to construct an affinity graph and track the evolution of the affinity between individuals over time, in order to predict affinity relationships arising through the influence of certain individuals within an online community. One of the added values of HAR-search is its ability to follow the temporal evolution of affinity relationships. HAR-search investigates the evolution of relationships between individuals through their affinity score by examining whether this score has remained constant, increased or decreased at any given time.

An affinity graph, $\mathcal{G=(U,E)}$, is a weighted graph where each node $u\in \mathcal{U}$ represents a user, the edge ($u,v$) $\in \mathcal{E}$ denotes an affinity relation between users $u$ and $v$, and the weight $w_{uv}\in {\rm I\!R}$ depicts the affinity score between the two users. If edge ($u,v$) does not exist, then the value of $w_{uv}$ is equal to 0. In this paper, we keep only edges for which $w_{uv}$ is greater than or equal to $10^{-5}$. An affinity graph is symmetric if and only if $w_{uv}=w_{vu}$, for all $(u,v), (v,u)\in \mathcal{E}$.

Since the focus in this paper is on investigating the influence of personality on affinity formation, we refer to the affinity graph as a triplet $\mathcal{G^{'}=(U,E,P)}$, where $\mathcal{P}=\{p_1,\dots,p_n\}$ represents the $16$ MBTI personality types, $\mathcal{U}$ is a finite node set that includes $n$ labels, $\mathcal{E}\subseteq\mathcal{U}\times\mathcal{U}$ is an edge set and $w_{uv}$ denotes the weight on edge $(u,v)$. We assign to each node $u$ a label corresponding to an MBTI personality type, $p_i\in\mathcal{P}$. Note that each node $u$ in $\mathcal{G^{'}}$ possesses only a single MBTI personality type $p_i$. Formally, $\mathcal{L}$ is a mapping function for labeling nodes in $\mathcal{G^{'}}$, $\mathcal{L} :\mathcal{U\xrightarrow{}P}$ such that $\mathcal{L}(u) = p_i$ is the label for node $u$. We assume that each node is associated with a given label in $\mathcal{P}$.

\subsection{Detecting the Influence of Personality on Affinity}\label{DetectingInflPErsAffinity}

To discover the influence of personality on affinity, we propose to cluster the nodes of graph $\mathcal{G^{'}}$ into groups of densely connected regions based on edge weights, i.e., affinity scores. To partition graph $\mathcal{G^{'}}$ into $k$ overlapping clusters such that $C_1\cup\dots\cup C_k\subseteq \mathcal{U}$, we use two different graph clustering techniques: the random walk hitting time-based digraph clustering algorithm K-destinations \cite{paXXX} and the Markov cluster algorithm MCL \cite{paMCLUn}. The rationale for using these algorithms in an affinity context is that they are based on first-order Markov chains, deal with directed graphs, and draw on  intuition from random walks on graphs to detect cluster structure. 

{\color{black}To determine the influence of personality on affinity, we count the number of links that each node has in a cluster $C_i$\iffalse without regarding the amount of affinity score\fi. Since each node is labeled with an MBTI personality type, we seek to discover nodes \iffalse of particular personality type\fi that include more links with nodes of different personality types within a cluster. We apply the same logic to all clusters to investigate the possible influence of personality on affinity. We assume that the overall number of links related to a specific personality type within a cluster demonstrates its openness to other types and this can be considered as a relevant signal of influence.}

Here, we describe the functioning of MCL and K-destinations. MCL proposes the following intuition for the graph clustering paradigm: $(i)$ the number of higher–length paths in $\mathcal{G^{'}}$ is large for pairs of nodes lying in the same dense cluster and small for pairs of nodes belonging to different clusters. $(ii)$ A random walk in $\mathcal{G^{'}}$ that visits a dense cluster will likely not leave the cluster until many of its nodes have been visited. $(iii)$ Considering all shortest paths between all pairs of nodes of $\mathcal{G^{'}}$, edges between different dense clusters are likely to be in many shortest paths. Specifically, MCL operates on a graph where the edge weights in the graph represent similarity scores and relies on the observation that a random walk is more likely to stay in a cluster rather than travel across the clusters. MCL iteratively alternates between two successive steps of expansion and inflation until it converges. The expansion step performs random walks of higher lengths and it enables connecting to different regions in the graph. The inflation step aims to strengthen the intra-cluster connections and weaken the inter-cluster connections \cite{paMCLUn,paMCLDeux}.

K-destinations is an iterative clustering algorithm which uses the asymmetric pairwise measure of Markov random walk hitting time on directed graphs to cluster the data. K-destinations partitions the nodes of the directed graph into disjoint sets using the local distribution information of the data and the global structural information of the directed graph. Specifically, K-destinations suggests the following steps for graph clustering. $(a)$ K-destinations initially fixes the destination nodes and assigns each sample to the cluster that has minimal hitting time from it to the destination node corresponding to the cluster. $(b)$ Then, in each cluster, K-destinations updates the destination node from the samples that minimize the sum of the hitting times from all samples in the cluster to the destination node. The clustering algorithm repeats the two steps $(a)$ and $(b)$ until the cluster membership of each sample does not change \cite{paXXX}.

\begin{table*}[t!]
\fontsize{9pt}{9pt}\selectfont
\setlength\tabcolsep{2.4pt}
\centering
\caption{Semantic similarity for affinity relationships between different MBTI personality types. Bold font indicates similarity scores greater than or equal to {0.2}. }\label{tabAffPersSemSimCosDistance0001}
\begin{tabular}{lcccccccccccccccc}
\toprule
      & ISTJ  & ISFP  & INFP  & ESFJ  & ISTP  & ISFJ  & INFJ  & ENTP  & INTP  & INTJ  & ESFP  & ENTJ  & ESTP  & ESTJ  & ENFP  & ENFJ \\ \toprule
ISTJ&--&--&--&--&--&--&--&--&--&--&--&--&--&--&--&--\\
ISFP&{\bf0.207}&--&--&--&--&--&--&--&--&--&--&--&--&--&--&--\\
INFP&0.002&{\bf0.311}&--&--&--&--&--&--&--&--&--&--&--&--&--&--\\
ESFJ&0.009&{\bf0.301}&0.03&--&--&--&--&--&--&--&--&--&--&--&--&--\\
ISTP&{\bf0.214}&{\bf0.208}&0.005&0.01&--&--&--&--&--&--&--&--&--&--&--&--\\
ISFJ&{\bf0.217}&{\bf0.21}&0.027&{\bf0.245}&0.142&--&--&--&--&--&--&--&--&--&--&--\\
INFJ&0.13&0.111&{\bf0.259}&0.102&0.076&{\bf0.218}&--&--&--&--&--&--&--&--&--&--\\
ENTP&0.005&0.007&0.082&0.086&0.115&0.033&0.011&--&--&--&--&--&--&--&--&--\\
INTP&0.08&0.1&{\bf0.304}&0.002&{\bf0.243}&0.007&0.094&{\bf0.232}&--&--&--&--&--&--&--&--\\
INTJ&{\bf0.226}&0.006&0.188&0.039&0.108&0.015&{\bf0.222}&0.101&{\bf0.209}&--&--&--&--&--&--&--\\
ESFP&0.04&{\bf0.209}&0.076&{\bf0.227}&0.062&0.11&0.003&0.095&0.082&0.003&--&--&--&--&--&--\\
ENTJ&0.1&0.004&0.003&{\bf0.201}&0.007&0.002&0.071&{\bf0.307}&0.074&{\bf0.2}&0.046&--&--&--&--&--\\
ESTP&0.098&0.104&0.005&0.143&0.199&0.002&0.005&0.186&0.01&0.004&0.178&0.113&--&--&--&--\\
ESTJ&{\bf0.217}&0.06&0.001&{\bf0.235}&0.105&0.076&0.026&0.113&0.008&0.057&0.105&{\bf0.252}&{\bf0.264}&--&--&--\\
ENFP&0.001&0.097&0.108&0.118&0.027&0.001&0.041&{\bf0.264}&0.005&0.004&{\bf0.301}&0.108&0.105&0.026&--&--\\
ENFJ&0.002&0.006&0.101&{\bf0.253}&0.001&0.104&0.06&0.119&0.013&0.025&0.114&{\bf0.257}&0.061&0.114&{\bf0.289}&--\\

\bottomrule
\end{tabular}
\end{table*}

\section{Experiments}\label{PrsAffExp43tGhZcqk73}
In this section, we show that the content of interactions between individuals self-identified with an MBTI type can be used to discover affinity relationships and analyze semantic similarity and emotional stability between affinity pairings. We further predict the MBTI personality types.\\

\noindent
\textbf{Affinity discovery.} In order to measure affinity relationships between individuals, we apply the HAR-search method to empirically quantify affinity connections in Dataset B. The results reported in Figure \ref{FigXDESQIO7865} show that affinity relationships ESFJ-ISFP and INTP-INFJ achieved the highest percentages (1.86\% and 1.58\%, respectively), and the affinity relationship ISFJ-ISFJ (1.51\%) is the only relationship between individuals of the same personality type that reports such a high percentage. Crucially, we observe that ESFJ-ISFP and INTP-INFJ also have relatively high semantic similarity scores (Table \ref{tabAffPersSemSimCosDistance0001}) and low Pearson correlation coefficients for negative emotions (Tables \ref{tabERECCCCC} and \ref{tabXD72iC}).

\begin{equation}
\label{EqCoSiml222}
\fontsize{9pt}{9pt}\selectfont
\cos ({ D_i},{ D_j})= {{ D_i} { D_j} \over \|{ D_i}\| \|{ D_j}\|} = \frac{\sum\limits_{i=1}^{n}{{ a}_i{ b}_j} }{\sqrt{\sum\limits_{i=1}^{n}{{ a}_i^2}} \sqrt{\sum\limits_{j=1}^{n}{{ b}_j^2}} }
\end{equation}

\noindent
\textbf{Semantic similarity.} To measure semantic similarity at the personality level, we regard only affinity relationships composed of people from two different MBTI personality types. We take all tweets belonging to people of the same personality type and assemble them in a single document (corpus). In total, we obtain 16 documents, $\{\text{D}_1,\dots,\text{D}_m\}$, $m=$16. We utilize GloVe word embedding \cite{GloveWE}, an unsupervised learning algorithm for obtaining vector representations for words in each document $\text{D}_i$. Specifically, $\text{D}_i=\{a_1, a_2, \dots\}$ and $\text{D}_j=\{b_1, b_2, \dots\}$ denote the vector representations of two different documents, for example {ISTJ} and {ISFP}. We use cosine similarity (see Eq. (\ref{EqCoSiml222})) to compute the semantic similarity of words for two documents $\text{D}_i$ and $\text{D}_j$ using their vector representations. The cosine similarity is measured by the cosine of the angle between two vectors and determines whether two vectors are pointing in approximately the same direction.

Table \ref{tabAffPersSemSimCosDistance0001} shows semantic similarity scores for affinity pairings composed of different MBTI personality types. To understand affinity formation, we investigate the semantic similarity scores more deeply from a personality standpoint. To this end, we regard arbitrarily the threshold for affinity relationships for which similarity scores are greater than or equal to {0.2}:
ISFP-ISTJ (0.207), ISTP-ISTJ (0.214), ISFJ-ISTJ (0.217), INTJ-ISTJ (0.226), ESTJ-ISTJ (0.217), INFP-ISFP (0.311), ESFJ-ISFP (0.301), ISTP-ISFP (0.208), ISFJ-ISFP (0.21), ESFP-ISFP (0.209), INFJ-INFP (0.259), INTP-INFP (0.304), ISFJ-ESFJ (0.245), ESFP-ESFJ (0.227), ENTJ-ESFJ (0.201), ESTJ-ESFJ (0.235), ENFJ-ESFJ (0.253), INTP-ISTP (0.243), INFJ-ISFJ (0.218), INTJ-INFJ (0.222), INTP-ENTP (0.232), ENTJ-ENTP (0.307), ENFP-ENTP (0.264), INTJ-INTP (0.209), ENTJ-INTJ (0.2), ENFP-ESFP (0.301), ESTJ-ENTJ (0.252), ENFJ-ENTJ (0.257), ESTJ-ESTP (0.264) and ENFJ-ENFP (0.289). Based on the preceding, it can be seen that ENFJ, ENFP, INFP, ENTJ, ENTP, ESFJ, ESTJ, and INTP each appear in two or three affinity relationships for which the similarity scores are superior to {0.23}. Specifically, the types ESFJ, ENTJ, and ENFP have the highest semantic similarity scores with ENFJ. Moreover, we note a number of affinity relationships with low semantic similarity scores: INFP-ISTJ (0.002), ENFP-ISTJ (0.001), ENFJ-ISTJ (0.002), ESTJ-INFP (0.001), INTP-ESFJ (0.002), ENFJ-ISTP (0.001), ENFP-ISFJ (0.001), ESFP-INFJ (0.003) and ESFP-INTJ (0.003).

\begin{table*}[t!]
\fontsize{9pt}{9pt}\selectfont
\setlength\tabcolsep{2.4pt}
\centering
\caption{Pearson correlations between LIWC (positive emotions) features extracted on language use to discover emotional stability in affinities between two different personality types. All correlations are significant at $p<0.01$.}\label{tabERECCCCC}
\begin{tabular}{lcccccccccccccccc}
\toprule
      & ISTJ  & ISFP  & INFP  & ESFJ  & ISTP  & ISFJ  & INFJ  & ENTP  & INTP  & INTJ  & ESFP  & ENTJ  & ESTP  & ESTJ  & ENFP  & ENFJ \\ \toprule
ISTJ&{--}&{--}&{--}&{--}&{--}&{--}&{--}&{--}&{--}&{--}&{--}&{--}&{--}&{--}&{--}&{--}\\

ISFP&0.041&{--}&{--}&{--}&{--}&{--}&{--}&{--}&{--}&{--}&{--}&{--}&{--}&{--}&{--}&{--}\\

INFP&0.222&0.046&{--}&{--}&{--}&{--}&{--}&{--}&{--}&{--}&{--}&{--}&{--}&{--}&{--}&{--}\\

ESFJ&0.065&0.272&0.054&{--}&{--}&{--}&{--}&{--}&{--}&{--}&{--}&{--}&{--}&{--}&{--}&{--}\\

ISTP&0.249&0.114&0.041&0.107&{--}&{--}&{--}&{--}&{--}&{--}&{--}&{--}&{--}&{--}&{--}&{--}\\

ISFJ&0.207&0.201&0.103&0.214&0.276&{--}&{--}&{--}&{--}&{--}&{--}&{--}&{--}&{--}&{--}&{--}\\

INFJ&0.113&0.011&0.070&0.106&0.204&0.212&{--}&{--}&{--}&{--}&{--}&{--}&{--}&{--}&{--}&{--}\\

ENTP&0.302&0.300&0.296&0.219&0.315&0.283&0.270&{--}&{--}&{--}&{--}&{--}&{--}&{--}&{--}&{--}\\

INTP&0.260&0.285&0.248&0.315&0.297&0.250&0.268&0.313&{--}&{--}&{--}&{--}&{--}&{--}&{--}&{--}\\

INTJ&0.318&0.209&0.287&0.107&0.288&0.265&0.313&0.285&0.322&{--}&{--}&{--}&{--}&{--}&{--}&{--}\\

ESFP&0.066&0.058&0.032&0.043&0.174&0.109&0.047&0.041&0.041&0.197&{--}&{--}&{--}&{--}&{--}&{--}\\

ENTJ&0.134&0.201&0.079&0.176&0.256&0.314&0.311&0.309&0.295&0.320&0.102&{--}&{--}&{--}&{--}&{--}\\

ESTP&0.217&0.123&0.009&0.150&0.162&0.091&0.079&0.252&0.173&0.088&0.076&0.222&{--}&{--}&{--}&{--}\\

ESTJ&0.311&0.308&0.325&0.238&0.249&0.248&0.254&0.238&0.038&0.024&0.104&0.274&0.215&{--}&{--}&{--}\\

ENFP&0.036&0.106&0.022&0.129&0.151&0.056&0.116&0.214&0.106&0.017&0.028&0.127&0.109&0.212&{--}&{--}\\

ENFJ&0.205&0.223&0.087&0.183&0.295&0.237&0.202&0.207&0.209&0.195&0.170&0.251&0.203&0.194&0.083&{--}\\ \bottomrule

\end{tabular}
\end{table*}

\begin{table*}
\fontsize{9pt}{9pt}\selectfont
\setlength\tabcolsep{2.4pt}
\centering
\caption{Pearson correlations between LIWC (negative emotions) features extracted on language use to discover emotional stability in affinities between two different personality types. All correlations  are significant at $p<0.01$.}\label{tabXD72iC}
\begin{tabular}{lcccccccccccccccc}
\toprule
      & ISTJ  & ISFP  & INFP  & ESFJ  & ISTP  & ISFJ  & INFJ  & ENTP  & INTP  & INTJ  & ESFP  & ENTJ  & ESTP  & ESTJ  & ENFP  & ENFJ \\ \toprule
ISTJ&{--}&{--}&{--}&{--}&{--}&{--}&{--}&{--}&{--}&{--}&{--}&{--}&{--}&{--}&{--}&{--}\\
ISFP&0.093&{--}&{--}&{--}&{--}&{--}&{--}&{--}&{--}&{--}&{--}&{--}&{--}&{--}&{--}&{--}\\
INFP&0.001&0.215&{--}&{--}&{--}&{--}&{--}&{--}&{--}&{--}&{--}&{--}&{--}&{--}&{--}&{--}\\

ESFJ&0.007&0.003&0.086&{--}&{--}&{--}&{--}&{--}&{--}&{--}&{--}&{--}&{--}&{--}&{--}&{--}\\

ISTP&0.002&0.107&0.013&0.022&{--}&{--}&{--}&{--}&{--}&{--}&{--}&{--}&{--}&{--}&{--}&{--}\\

ISFJ&0.001&0.083&0.025&0.008&0.021&{--}&{--}&{--}&{--}&{--}&{--}&{--}&{--}&{--}&{--}&{--}\\

INFJ&0.085&0.236&0.114&0.013&0.017&0.004&{--}&{--}&{--}&{--}&{--}&{--}&{--}&{--}&{--}&{--}\\

ENTP&0.002&0.003&0.001&0.007&0.001&0.001&0.002&{--}&{--}&{--}&{--}&{--}&{--}&{--}&{--}&{--}\\

INTP&0.001&0.002&0.003&0.002&0.001&0.002&0.003&0.001&{--}&{--}&{--}&{--}&{--}&{--}&{--}&{--}\\

INTJ&0.001&0.007&0.002&0.003&0.002&0.001&0.001&0.002&0.001&{--}&{--}&{--}&{--}&{--}&{--}&{--}\\

ESFP&0.074&0.119&0.068&0.209&0.019&0.026&0.034&0.037&0.075&0.013&{--}&{--}&{--}&{--}&{--}&{--}\\

ENTJ&0.001&0.002&0.002&0.003&0.001&0.002&0.001&0.001&0.002&0.001&0.009&{--}&{--}&{--}&{--}&{--}\\

ESTP&0.010&0.048&0.207&0.027&0.034&0.077&0.018&0.004&0.008&0.013&0.025&0.008&{--}&{--}&{--}&{--}\\

ESTJ&0.001&0.001&0.003&0.009&0.005&0.003&0.004&0.016&0.004&0.036&0.031&0.004&0.009&{--}&{--}&{--}\\

ENFP&0.206&0.004&0.116&0.118&0.023&0.115&0.015&0.021&0.013&0.034&0.116&0.012&0.016&0.007&{--}&{--}\\

ENFJ&0.003&0.002&0.019&0.015&0.004&0.008&0.006&0.013&0.007&0.008&0.083&0.001&0.205&0.009&0.011&{--}\\ 

\bottomrule
\end{tabular}
\end{table*}

\noindent
\textbf{Emotional stability.} To measure emotional stability in affinity relationships between two different personality types, we investigate language use in their discussion interactions and utilize the Linguistic Inquiry and Word Count (LIWC) text-analysis system to extract psycholinguistic features. LIWC is a widely used, psychometrically validated system for psychology-related language analysis and word classification. The LIWC dictionary includes word categories that have pre-labeled meanings created by psychologists. The LIWC categories have also been independently evaluated for their correlation with psychological concepts \cite{Pennebaker:15}. For each tweet, we computed the number of observed words and terms using the LIWC system and focusing exclusively on two LIWC categories: psychological processes and linguistic dimensions. For the psychological processes, we focused especially on the following two subcategories: positive and negative emotions. With regard to the linguistic dimensions category, we measured solely the proportion of first-person pronouns in the tweet content. Research shows that pronouns reveal information on a person's emotional state, thinking, and personality \cite{Pennebaker:15}.  \cite{Chung:07} found that individuals who are strongly susceptible to emotional reactions or vulnerable situations more frequently use first-person pronouns, suggesting higher self-attention focus.\\

\noindent
\textbf{Pearson correlations from LIWC features.} We performed linear regression with elastic-net regularization to calculate Pearson correlation coefficients, using the weights of the LIWC features. Let $X=\{x_1,x_2,\dots\}$ and $Y=\{y_1,y_2,\dots\}$ denote two feature vectors extracted from the language use of two different personality types. To compute Pearson's {\it r}, we took the top $n$ elements of each vector in descending order ($n=$1000); the complete results can be seen in Tables \ref{tabERECCCCC} and \ref{tabXD72iC}. 

Tieger and Barron-Tieger~\shortcite{paTieger} explored the personality type of many couples and found that the more type preferences a couple had in common, the more satisfied they were with their communication. In the work reported here, we found that the personality types bearing the preferences {{S}} (sensing) and {{J}} (judgment), that is, {{ESTJ}, {ESFJ}, {ISTJ}} and {{ISFJ}}, are not emotional when they are in affinity relationships among themselves. We also found that {{ENTP}, {INTP}, {INTJ}, {ENTJ}} and {ESTJ} maintain good affinity relationships with all personality types and tend to be emotionally stable people (Tables \ref{tabERECCCCC} and \ref{tabXD72iC}). Our results support the outcomes of the study conducted by \cite{paTieger} on couples and personality type, except for {{ENTP}, {INTP}, {INTJ}, {ENTJ}} and {ESTJ}.  Tieger and Barron-Tieger's research found that $(i)$ {{ESTJ}, {ESFJ}, {ISTJ}} and {ISFJ} have a satisfaction rate of 79\% when paired with each other, and $(ii)$ {{ENFP}, {INFP}, {ENFJ}} and {INFJ} have a satisfaction rate of 73\% when paired with each other. These tend to place a high value on relationships and are the most likely of all the types to devote themselves to healthy relationships and open communication. \\

\begin{table}[t!]
\fontsize{10pt}{10pt}\selectfont
\setlength\tabcolsep{2.4pt}
\centering
\caption{Prediction results of MBTI personality types. Bold font indicates the best performance for each MBTI type. }
\label{tabXVC45aYTDe}
\begin{tabular}{lccccc}
\toprule
      & LR    & RF    & SVM   & NB  & BERT \\ \toprule
ISTJ  & 0.753 & 0.782 & 0.786 & 0.721 & {\bf0.891}  \\
ISFP  & 0.761 & {\bf0.777} & 0.766 & 0.711 & 0.773  \\
INFP  & 0.774 & {\bf0.802} & 0.774 & 0.698 & 0.800  \\
ESFJ  & 0.780 & 0.783 & 0.776 & {\bf0.763} & 0.785  \\ 
ISTP  & 0.782 & 0.769 & 0.782 & 0.735 & {\bf0.888}  \\
ISFJ  & 0.755 & 0.757 & 0.739 & 0.697 & {\bf0.812}  \\
INFJ  & {\bf0.775} & 0.768 & 0.757 & 0.747 & 0.774  \\
ENTP  & 0.779 & 0.773 & {\bf0.782} & 0.709 & 0.781  \\
INTP  & 0.754 & 0.756 & 0.745 & 0.705 & {\bf0.859}  \\
INTJ  & 0.798 & 0.785 & 0.795 & 0.722 & {\bf0.803}  \\
ESFP  & 0.759 & 0.760 & 0.757 & 0.698 & {\bf0.866}  \\
ENTJ  & 0.781 & 0.778 & 0.769 & 0.767 & {\bf0.887}  \\
ESTP  & 0.758 & 0.757 & {\bf0.763} & 0.760 & 0.762  \\
ESTJ  & 0.784 & 0.789 & 0.773 & 0.755 & {\bf0.894}  \\
ENFP  & 0.780 & 0.766 & 0.782 & 0.699 & {\bf0.806}  \\
ENFJ  & 0.767 & 0.782 & 0.774 & 0.731 & {\bf0.861}  \\  \bottomrule
\end{tabular}
\end{table}

\noindent
\textbf{MBTI prediction.} 
In order to predict each of the 16 MBTI personality types, we trained five different classifiers: logistic regression (LR, $10^8$ ridge), random forest (RF) with AdaBoost, support vector machine (SVM), a simple naive Bayes (NB) and BERT \cite{b0000022}. For SVM, we set the regularization parameter $\lambda$ to 0.0001  and the value $\gamma$ of the radial basis function kernel to 0.5; for RF, we set the number of trees to 500 and the maximum depth and number of features to 3 and 30, respectively. For BERT, we used the BERT-large-cased model, which comprises 24 layers, 16 attention heads and 340 million parameters. We conducted multi-class classification by extracting and analyzing linguistic patterns from the user tweets and personality labels mentioned in Dataset B (see Section \ref{PrsAffDtsHvSkO35ZWp82n}).

To evaluate the performance of the constructed multi-class classifiers, we performed 10-fold cross-validation to split our training and testing sets and computed the F-1 score metric to measure the accuracy of our classifiers. Table \ref{tabXVC45aYTDe} presents the performance results of the five classifiers. We report that the F-1 scores for our classifiers are relatively highs and show the ability to predict all of the 16 MBTI personality types. It can be observed that the majority of the best performances were achieved by BERT, with F-1 scores of over 0.8. Even BERT's poorer results outperforms some of the other classifiers by a significant margin. Interestingly, we found that ESTJ, ENTJ and ISTP were easily predicted by the five classifiers utilized, as they yielded the highest average performances: 0.799, 0.796 and 0.791, respectively. In particular, it can be seen that the personality types containing the preferences T (thinking) and J (judgment) yielded higher average performance. We also note that RF consistently performed well on the personality types bearing the preferences I (introversion), F (feeling)  and P (perception), and SVM surpassed all classifiers on the personality types that include the preferences E (extraversion), T (thinking), and P (perception). \\

\begin{table}[t!]
\fontsize{8.8pt}{8.8pt}\selectfont
\setlength\tabcolsep{2.1pt}
\centering
\caption{Clustering results in terms of Error and NMI. Bold font indicates the best performances. }\label{tabXVCVIClusterPerformance}
\begin{tabular}{lcccc} \toprule
      & MCL & 5-destinations & 10-destinations & 15-destinations \\ \toprule
NMI   & 0.822 & {\bf0.848} & 0.814 & 0.797  \\ \toprule
Error & 0.016 & {\bf0.013} & 0.030 & 0.071 \\ \bottomrule
\end{tabular}
\end{table}

\noindent
\textbf{Influence of personality on affinity.} To discover the influence that the personality types have on affinity formation, we utilized the approach proposed in Section \ref{DetectingInflPErsAffinity}. MCL is an unsupervised graph clustering algorithm.. For K-destinations, we set the number of destination nodes by varying K, assigning it values of 5, 10 and 15. This variation allows us to better explore the influential personality types on various facets. 

To evaluate the performances of MCL and K-destinations, we computed two performance measures from the clustering results: the normalized mutual information (NMI) and the minimal clustering error (Error). The NMI is defined as

\begin{equation}
\label{EqCoSiml22244}
\fontsize{9pt}{9pt}\selectfont
\text{NMI}= \frac{I(x,y)}{\sqrt{H(x)H(y)}}, 
\end{equation}

\noindent
where $I(x,y)$ is the mutual information between the true $x$ and $y$, and $H(x)$ and $H(y)$ are the entropies of $x$ and $y$, respectively. Note that $0\leq\text{NMI}(x, y)\leq1$ and $\text{NMI}(x, y) = 1$ when $x = y$. The larger the value of NMI, the better the clustering result.

The clustering error is defined as the minimal classification error among all possible permutation mappings, defined as:

\begin{equation}
\label{EqCoSiml22233}
\fontsize{9pt}{9pt}\selectfont
\text{Error} = \min(1 - \frac{1}{n}\sum\limits_{i=1}^{n}{\delta(y_i - \text{perm}(c_i))}),
\end{equation}

\noindent
where $y_i$ and $c_i$ are the true class label and the obtained clustering result of $x_i$, respectively, and $\delta(x, y)$ is the delta function
that equals $1$ if $x = y$ and $0$ otherwise.

The clustering results for the two methods are summarized in Table \ref{tabXVCVIClusterPerformance}. Our results demonstrate that we achieved good performance for graph clustering. It can be seen that 5-destinations achieved significantly better performance on both evaluation metrics, that is, the smallest error and the largest NMI values. Moreover, we observe that the error values for K-destinations become significantly larger as the set value of K increases, showing that this variation can reduce the NMI value by a considerable margin. Specifically, we extracted 6, 4, 7 and 9 clusters with MCL, 5-destinations, 10-destinations and 15-destinations, respectively. As described in Section \ref{DetectingInflPErsAffinity}, we counted the number of links that each node has in each cluster. We assume that the total number of links in a set of nodes indicates a relevant signal of influence. Specifically, for each cluster, we report only the node with the highest number of links. We obtained (ENTJ, ENFP, ESTJ, INTP, ISTJ, INFP) for MCL, (ESTJ, INTP, ENFP, ENTJ) for 5-destinations, (INFP, ENFP, ISTJ, INTJ, ESTJ, ENTJ, INTP) for 10-destinations and (INFP, INTJ, INTP, ISTP, ENFP, ESTP, ENTJ, ISTJ, ESTJ) for 15-destinations. Note that the four influential personality types extracted from 5-destinations are also part of MCL, 10- and 15-destinations.

\section{Discussion}\label{PrDscss9189erhd091ue}
Our results provide some of the first insights into the investigation and understanding of affinity relationships between personality types on social media. We measured semantic similarity and emotional stability in affinities, and showed the feasibility of applying clustering to discover the influence of personality on affinity. Moreover, we trained five different classifiers from the spontaneous language utilized by a set of social media users to predict the 16 MBTI personality types. The theoretical and practical implications of our outcomes can be valuable for supporting decision-making processes in various domains, including clinical psychology, forensic psychology, digital forensics, human factors and social science.

Our results identify a number of statistically significant correlations in terms of emotional stability in personality-based affinity relationships. It should be recalled that our investigation was limited to extracting LIWC features and measuring correlation coefficients related to emotional stability in affinity pairings. This study does not specifically examine the reasons or the circumstances in which emotional reactions were expressed. Importantly, we report 13 affinity pairings for which correlation values for negative emotions surpassed 0.1: {{ENFP-ISTJ} (0.206)}, {{INFP-ISFP} (0.215)}, {{ISTP-ISFP} (0.107)}, {{INFJ-ISFP} (0.236)}, {{ESFP-ISFP} (0.119)}, {{INFJ-INFP} (0.114)}, {{ESTP-INFP} (0.207)}, {{ENFP-INFP} (0.116)}, {{ESFP-ESFJ} (0.209)}, {{ENFP-ESFJ} (0.118)}, {{ENFP-ISFJ} (0.115)}, {{ENFP-ESFP} (0.116)} and {{ENFJ-ESTP} (0.205)}. We note that only {{ENFP}} and {{ISFP}} appear in five and four different affinity pairings, respectively. Our findings show strong evidence that the types {{ENFP}} and {{ISFP}} are particularly emotionally reactive and predominantly mention negative emotions in their narratives. The two types appear quite close in terms of affinity percentage (1.38\%, see Figure \ref{FigXDESQIO7865}) and have in common two preferences ({{F}} and {{P}}). Moreover, we note that the 13 aforementioned affinity pairings have relatively high semantic similarity scores, except for {{ENFP-ISTJ}}, {{ESTP-INFP}}, {{ENFP-ISFJ}} and {{ENFJ-ESTP}}. From our experiments, our observation is that emotional stability does not depend strongly on semantic similarity. For instance, we find that affinity pairings with semantic similarity scores less than or equal to 0.003 have high and low correlation values for positive and negative emotions, respectively, except for {{ENFP-ISTJ}}, {{ENFP-ISFJ}} and {{ESFP-INFJ}}. We believe that the findings on semantic similarity and emotional stability constitute an important lead for understanding the implications of personality in the development of affinity.

An interesting thing to note about the cluster analysis is that our findings suggest the value of K can greatly affect the ability of K-destinations to accurately detect clusters in the affinity graph. We therefore explored the clusters detected by both MCL and K-destinations to extract influential personality types. Before proceeding further, it should be noted that we limited ourselves to identifying the influence of personality on affinity. Applying our approach to the results yielded by the clustering techniques used, we identify potential influential personality types for each cluster and observe that these influential personality types overlap from one technique to another. We remark that the four influential personality types stemming from 5-destinations can also be found in MCL, 10-destinations and 15-destinations. However, analyzing the aspects on which certain personality types were influenced by the influential MBTI types yielded by 5-destinations constitutes a challenging problem and naturally requires further inquiry. As mentioned earlier, most studies on personality rely more heavily on questionnaires to evaluate individual preferences and predict team dynamics \cite{paOmarTypes}. Combining social influence-based behavior questionnaires and social media interactions may possibly reveal important factors that can help investigate and explain the causes of influence with sufficient certainty. In reality, investigations into the influence of personality can be driven by the concrete needs of applications. Examples might be investigating the role that personality plays in the effective functioning of  behavioral deterioration \cite{paJMDeterioration}. Our results also contribute to understand affinity-seeking behaviors and affinity-maintaining patterns between relationships of individuals of different personality types. Our approach can be used as baseline to detect affinity-seeking behaviors from textual data stemming from social media.

\setlength{\parskip}{0pt}

Applying social media users' self-identified types and examining their spontaneous language, we extracted linguistic patterns using five different classifiers to predict the 16 MBTI personality types. The results are very encouraging and show that our classifiers can effectively predict personality with high accuracy. In particular, we achieved the majority of the best performances with BERT. BERT predicted the personality by not only considering self-reported type classes but also capturing the context in which the text corpus related to each type class was expressed. To validate the performance of the classifiers used, we considered self-identified types as ground truths. A major advantage of using the self-identified types as ground truths is their ability to act as immediate validation \cite{paFrommetPaper}. We recognize that an individual's personality could possibly develop and change over time \cite{paPersoOverTime}. To predict the personality types, we consistently pre-processed the experimental dataset to remove individuals who have reported two or more types. In the future, we would like to keep individuals with several self-identified types, in order to investigate the dynamic nature of personality. We believe that data processing can potentially contribute to the ever-challenging  task of personality prediction from social media text data.

\section{Conclusion}\label{PrsAffClsO73ByZol}
In this paper, we presented a series of analyses to understand affinity relationships between personality types on social media. Specifically, we focused strongly on individuals who self-identified with one of the MBTI types, and explicitly tracked their language use. Our results have shown significant correlations in emotional stability in affinity relationships between individuals from different personality types, and examined the semantic similarity in these affinity relationships. In addition to these analyses, we have provided new insights for discovering the influence that certain personality types have on others and predicting personality by utilizing the linguistic patterns extracted directly from spontaneous language. Our study contributes to the body of research on personality, with a new understanding of the implications of the influence that personality has on affinity relationships.

\setlength{\parskip}{0pt}

While the scope of our study is limited to understanding the influence of personality on affinity by utilizing psycholinguistic features, our findings point the way for future investigations of broader scope. For instance, in exploring the influence of personality on affinity, the socioeconomic status and demographic information of individuals could be considered. We believe that this may provide additional insights, allowing examination of more subtle details that could help to better explain the influence of personality. Future studies may juxtapose psycholinguistic and demographic features to explore different facets of the influence of personality. Moreover, we aim to utilize demographic features to measure the correlation between socioeconomic status and affinity relationships.


\begin{table*}[]
\setlength\tabcolsep{4pt}
\centering
\caption{Characteristics of the MBTI types. Source: The Myers-Briggs Company (\url{https://eu.themyersbriggs.com/}) }\label{MBTICharacteristicsAffPersXX2eXOiP}
\begin{tabular}{ll}
\toprule
       MBTI &  Characteristics \\ \toprule
ISTJ & People with ISTJ preferences are typically thorough, conscientious, \\
 & realistic but also systematic and reserved. \\
ISFP & ISFPs are cooperative, modest and adaptable and also gentle and \\
 & loyal. \\
INFP & INFPs are flexible, spontaneous as well as reflective and contained. \\
 &  They are also imaginative and developmental. \\
ESFJ & ESFJs are warm and appreciative as well as organized, outgoing  \\
 &  and supportive. They are also realistic and loyal. \\

ISTP & ISTPs are analytical, practical, realistic but also logical and \\ & adaptable.\\
ISFJ & ISFJs are organized, practical and patient, but also dependable \\
 &  and loyal. Furthermore ISFJs are patient and understanding.\\
INFJ & INFJs are compassionate, idealistic as well as imaginative and \\ & visionary. They are also sensitive and reserved. \\
ENTP & ENTPs are emergent, theoretical and flexible as well as  \\
& imaginative and challenging. \\
INTP & INTPs are independent and detached, who also tend to be    \\
 &  challenging and logical as well as skeptical and innovative. \\
INTJ & INTJs are strategic and conceptual as well as innovative,   \\
 &  independent and logical. They can also be demanding and \\
 & reflective.\\
ESFP & ESFPs are tolerant and spontaneous as well as playful   \\
 & and resourceful. They also tend to be friendly and enthusiastic. \\
ENTJ & ENTJs are structured and challenging, they also tend to be  \\ 
& strategic and questioning.\\
ESTP & ESTPs are analytical, outgoing and enthusiastic as well as logical  \\
 & and they tend to be observant and resourceful. \\
ESTJ & ESTJs are responsible and efficient but they can also be   \\
& assertive as well as logical and realistic. \\
ENFP & ENFPs are friendly and expressive as well as innovative and \\ &  energetic. \\
ENFJ & ENFJs are warm, collaborative and supportive, as well as friendly \\
& and organized. They also tend to be persuasive. \\ \bottomrule  

\end{tabular}
\end{table*}

\end{document}